# Gate-tunable large magnetoresistance in an all-semiconductor spin-transistor-like device

M. Oltscher, F. Eberle, T. Kuczmik, A. Bayer, D. Schuh, D. Bougeard, M. Ciorga[*] and D. Weiss

Institute for Experimental and Applied Physics, University of Regensburg, 93053 Regensburg, Germany

**A large spin-dependent and electric field-tunable magnetoresistance (MR) of a two-dimensional electron system (2DES) is a key ingredient for the realization of many novel concepts for spin-based electronic devices[1–3]. The low MR observed during the last decades in devices with lateral semiconducting (SC) transport channels[4] between ferromagnetic (FM) source (S) and drain (D) contacts has been the main obstacle for realizing spin field effect transistor (sFET) proposals[2,3]. Here, we show both, a large two terminal MR in lateral 2DES-based spin valve geometry, with up to 80% resistance change, and tunability of the MR by an electric gate. The large MR is due to finite electric field effects at the FM/SC interface, which boost spin-to-charge conversion[5–8] and thus enhance the MR. The gating scheme we use is based on switching between uni- and bi-directional spin diffusion, without resorting to spin-orbit coupling (SOC)[9–11]. It can be therefore employed also in materials with low SOC.**

Since the original sFET proposal[2], spin injection into nonmagnetic semiconductors and conversion of spin information into large electric signals has become one of the main challenges of spintronics[12]. Efficient spin injection from a ferromagnet (FM) into a semiconducting (SC) channel requires a tunnel barrier between the FM and the SC[13]. This does not guarantee, however, a large magnetoresistance ratio $MR = \Delta R/R^P$, with $\Delta R = R^{AP} - R^P$, and $R^{P(AP)}$ being the two-terminal (2T) resistance of the device (see Fig. 1), measured between the ferromagnetic contacts when the magnetization of the electrodes is either parallel (P) or antiparallel (AP) to each other. A sizeable MR, being indispensable for novel spin-transistor concepts[2,3], requires that the electrons' dwell time in the transport channel is much smaller than the corresponding spin relaxation time[4,14]. This condition is particularly difficult to satisfy in semiconductors, as a high interface resistance needed for efficient injection increases also the dwell time, and is the reason for small MR < 1% obtained so far for semiconductor channels[4] (see Supplementary information 1).

---

[*]Corresponding author: mariusz.ciorga@ur.de

One aspect that has not been fully exploited in semiconductor spin devices, though, is how an electric field in the channel and in the contacts affects the MR. Albeit the influence of electric fields on both spin transport and spin-to-charge conversion in semiconducting channels[5,15] have been discussed before, the experiments have been typically limited to cases involving only a single biased FM/SC junction[6–8,16]. The standard two-terminal configuration implies, however, that the two FM/SC contacts are biased in opposite directions. Therefore, a non-rectifying *I-V* characteristic of the contacts is needed to drive a sufficiently large current through the device, thus generating a sizeable spin accumulation in the channel and allowing to exploit electric field effects.

In Fig. 1 and Methods we present devices that meet these requirements. Ferromagnetic source and drain of the transistor-like structures consist of (Ga,Mn)As/GaAs spin Esaki diodes[17,18] with nearly linear *I-V* characteristic at low bias and very efficient spin injection from ferromagnetic (Ga,Mn)As into both, bulk[19] and 2D GaAs[20] channels (see Supplementary information 1). Using a two-dimensional electron system (2DES) as a transport channel allows for additional control of spin accumulation in the channel through electric gates via SOC or, as used here, to confine spins in the region between the leads. This provides an extra knob for controlling the resistance of the device.

First, we present in Fig. 2 the large MR signals measured in a non-gated device. Fig. 2a shows corresponding *I-V* curves for P (Fig. 1c) and AP (Fig. 1d) configurations. The curves are almost symmetric with respect to *V*=0, enabling us to pass currents up to 100 µA through the device. Clearly, for a given voltage *V* the current is lower for the AP configuration, indicating a larger resistance due to an increased spin accumulation in the channel as sketched in Figs. 1c,d. The splitting between both curves increases with *V*, causing a larger MR for higher bias. This is confirmed by corresponding measurements of the spin valve (SV) effect, where the magnetic field *B* is swept along the contacts' axis (see Methods). Figs. 2b and 2c show typical SV signals measured in the linear (b) and in the nonlinear (c) regime of the Esaki diode. The full bias dependence of the MR is plotted in Fig. 2d. In the linear regime, MR = $\Delta R/R^P$ increases with bias, reaching ~20%. In the nonlinear regime, the signal is further enhanced when biased with constant current[7]. Then the difference between the *I-V* characteristic's P and AP branch leads to a large voltage output $\Delta V = V^{AP} - V^P$. With $\Delta V \cong 350$ mV measured for *I*=+66 µA this results in a MR ratio of ~80%. A similar behavior was observed on eight different samples. The corresponding data for one of the gated devices is exemplarily shown in Supplementary Information 2.

To demonstrate the key role of electric fields in obtaining large MR, we compare our $\Delta R$ data with values expected from the commonly accepted model of spin injection[13,21]. Following this model, a charge current *I* flowing across a FM/SC junction generates a spin accumulation $\mu^s$ in the SC channel. On the other hand, the spin accumulation in the vicinity of such a ferromagnetic contact leads to a spin-dependent voltage that can be measured due to Silsbee-Johnson spin charge coupling[22]. In 2T configuration, where charge current flows between source and drain, each FM contact serves both as injector and detector, i.e., a spin accumulation is created and detected at both contacts. (see Figs. 1c, d). The voltage drop at each FM/SC interface contains therefore two spin-dependent contributions, one from $\mu^s$ generated at this particular contact and a second one from the spin accumulation generated at the other contact. In AP configuration these voltages add up, whereas in P configuration they partially cancel out. As a result, the voltage difference between P and AP configuration, i.e. the total SV signal of the 2T device, is given by[23] $\Delta V = 2\delta V_{S,D} + 2\delta V_{D,S}$, where $\delta V_{i,j}$ is the contribution of the spin accumulation generated at contact $j$ that diffused to contact $i$. The different contributions to the SV signal $\Delta V$ are illustrated in Figs. 1c,d. $2\delta V_{i,j}$ corresponds directly to the SV signal $\Delta V_{i,j}^{nl}$ measured in the standard nonlocal configuration (Fig. 1b) with contact $j$ as injector and $i$ as detector. Hence, in the limit of low electric field ($E \cong 0$) the 2T SV signal can be approximated by $\Delta V^{E=0} = \Delta V_{S,D}^{nl} + \Delta V_{D,S}^{nl}$.

In Fig. 3 we compare the measured 2T resistance $\Delta R = \Delta V/I$ (black triangles) with $\Delta R^{E=0} = \Delta V^{E=0}/I$ (blue squares) obtained from corresponding nonlocal measurements at four injection currents in the linear regime (see also Supplementary Information 3). At a low bias of 0.7 µA, $\Delta R \cong \Delta R^{E=0}$, as expected for negligible electric fields. For higher currents, however, $\Delta R$ substantially exceeds $\Delta R^{E=0}$. The discrepancy increases with bias and at $I$=52 µA the measured $\Delta R$ is 32 times larger than the zero-field value $\Delta R^{E=0}$. This comparison demonstrates that the electric field is indeed responsible for the large MR ratios.

The electric field can affect MR in two ways. It can (i) alter the transport of $\mu^s$ along the channel[15] or (ii) influence the spin-to-charge conversion at the FM/SC junction[5–7]. We find below that the second effect is dominating here. It has been shown previously that the electric field of a positively biased FM/SC tunnel contact can significantly enhance the detected spin signal[5–8]. Here, for positively biased source and negatively biased drain, this means that $\Delta R_{S,D}$ is enhanced and $\Delta R_{D,S}$ suppressed compared to the low-bias values. From separate nonlocal experiments (Supplementary Information 3) we find that $\Delta R_{S,D}$ is enhanced by a factor of 4, 30 and 47 for $I$=10, 30 and 52 µA, respectively, whereas $\Delta R_{D,S}$ is either unchanged or slightly suppressed.

We also took effect (i) into account by calculating how drift enhances (suppresses) diffusion of spins from the drain (source) to the source (drain). The effect of drift on diffusion can be described by modified values of diffusion lengths, given by $\lambda_{s,d(u)} = \left[ -(+)\frac{|E|}{2}\frac{\mu}{D_s} + \sqrt{\left(\frac{|E|}{2}\frac{\mu}{D_s}\right)^2 + \frac{1}{\lambda_s^2}} \right]^{-1}$, where $\lambda_s$ is the intrinsic, zero-drift value and $\lambda_{s,d(u)}$ corresponds to the down-stream (up-stream) spin diffusion length in an electric field $E^{15}$. Here, $\mu$ is the electron mobility and $D_s$ is the spin diffusivity. Taking the corresponding values from experiment we find that for $I$=+52 µA the drift in the channel leads to an enhancement of $\Delta R_{S,D}$ by ~1.4 and a suppression of $\Delta R_{D,S}$ by a factor of two. Therefore, drift along the channel alone has only minor effect on the two-terminal spin signal. Considering both effects, (i) and (ii), we deduce (see Supplementary Information 3) a value for finite electric field $\Delta R^{E\neq 0}$ (red dots), which is in good agreement with the measured value $\Delta R$, as shown in Fig. 3.

Having demonstrated the drastic enhancement of the 2T MR by increasing the spin-to-charge conversion in our devices, we now address the most desired functionality of a spintronic device, i.e., efficient electrical control of $R$ via control of the spin state. In the seminal proposal of the sFET[2] and in related experiments[9–11,24] spin-orbit fields mediate sFET action. The realization of this concept requires materials with a spin-orbit coupling large enough for efficient rotation of travelling spins but small enough to prevent spin relaxation between S and D. For low SOC, e.g., the corresponding spin precession frequency is low what in turn calls for long transport channels. Therefore, also alternative approaches have been suggested, where transistor action is obtained either through electrical control of the spin injection process[3] or of the spin transport efficiency along the channel[25,26].

Here, we show a novel scheme of efficiently tuning $R$ without invoking SOC. The scheme is based on the control of the diffusion direction of the generated spin accumulation. Blocking diffusion of spins in the directions away from source and drain by gates (see sketch in Fig. 4a) and thus confining the spin accumulation to the region between S and D should result in larger spin signals compared to an open situation where spins are free to diffuse away from the contacts in both directions[23,27,28]. Employing gates allows us to switch in a controlled way between open and confined configuration, by depleting parts of the 2DES (see Fig. 4a). We indeed clearly observe the expected enhancement of the 2T spin valve signal as displayed in Fig. 4b. It is obtained for a bias current of 25 µA when switched from open to closed configuration. For this current, the MR is increased roughly six times from 3% to 18%. The observed enhancement $\Delta R_{conf}/\Delta R_{open}$ is very close to the value of 5.8 estimated for our device, based on the standard equations[23], in the limit of low electric fields (Supplementary Information 4). For increased bias values we however find the effect of confinement to be reduced. Similar results were also obtained for the second investigated device (Supplementary Information 5). Understanding of the exact mechanism behind this effect requires further investigations.

Combining the electric field driven boost of spin-to-charge conversion with a novel concept of gate control of spins, we have demonstrated a spin transistor-like device with at least one order of magnitude larger MR signals than reported so far for semiconductor spin valves. Harnessing the action of electric fields opens a pathway towards spin transistors with large MR, also for systems working up to room temperature[8,29,30]. The demonstrated spin transistor action, based on electrical control of spin diffusion in the channel, presents an attractive way of tuning the resistance of spin channels, particularly those defined in materials with low spin-orbit coupling.

**Methods**

**Fabrication.** All investigated samples were fabricated from the same wafer, grown epitaxially by molecular beam epitaxy. The wafer consists of the following layers (from top): ferromagnetic $Ga_{0.95}Mn_{0.05}As$ (50 nm), $Al_{0.33}Ga_{0.67}As$ (2 nm), $n^+$-GaAs (8 nm, $n^+= 5\times10^{18}$ cm$^{-3}$), $n^+\rightarrow n$-GaAs transition layer (15 nm,), $n$-GaAs (100 nm, $n= 6\times10^{16}$ cm$^{-3}$), i-GaAs (50 nm), $Al_{0.33}Ga_{0.67}As$ (125 nm), AlGaAs/GaAs superlattice buffer (500 nm) and (001)-oriented GaAs substrate. The top four layers form a spin Esaki diode. The two-dimensional electron system (2DES) is confined at the $i$-GaAs/$Al_{0.33}Ga_{0.67}As$ interface and charge carriers are provided by a Si δ-doped layer in the $Al_{0.33}Ga_{0.67}As$ region[20]. As the (Ga,Mn)As layer is degenerately $p$-doped, it is necessary to increase the Si doping in the $n^+$-layer to ensure Esaki diode functionality. This is achieved by so-called pseudo-δ-doping, meaning that the growth process of the 8 nm thick n+-GaAs layer using continuous Si flux is stopped every 1.6 nm for 10 s, to accumulate Si dopants. This allows n$^+$-doping higher than $1\times10^{19}$ cm$^{-3}$, making the $p$-$n$-junction more symmetric and thus lowering its resistance. 10 μm wide transport channels are defined using photolithography and wet etching techniques. Narrow, 500 and 700 nm wide, ferromagnetic electrodes are defined using electron beam lithography, followed by evaporation of Au/Ti contacts. Different widths of source and drain contacts ensure different switching fields of the contacts during magnetoresistance measurements. In the region between the contacts, the top 3 layers are partially etched away using wet chemical etching in order to limit the lateral transport to the 2DES. For the samples described in this paper, the etching depth was between 56 and 60 nm. The electric gates were defined using e-beam lithography followed by atomic layer deposition of 40 nm insulating $Al_2O_3$ and thermal evaporation of 7 nm NiCr. Such gates ensure ~70% transparency to red light (λ= 645 nm), used to illuminate the samples in order to populate the 2DES.

**Measurements.** Electrical measurements were performed primarily in the local configuration (see Fig. 1a), i.e. with the current flowing between the ferromagnetic source and drain contacts, either applying a constant current or a constant voltage. The voltage drop $V_{2T}$ is measured directly between both contacts. SV measurements were performed while sweeping an external magnetic field along the axes of the FM contact stripes between $B = 0.5$ T and $-0.5$ T, values needed to initially saturate the magnetization of the contacts along the external $\boldsymbol{B}$.

**Acknowledgements**

The work has been supported by the German Science Foundation via SFB689.

**Contributions**

All authors contributed to various aspects of the work presented here. M. O. and F. E. fabricated the devices and performed the measurements. T. K. assisted by device characterization measurements. M. O. and M. C. analyzed the data. A. B., D. S. and D. B. designed and realized the heterostructure. Project planning was done by M.C. and D.W.; M. C., M. O., D. W. and D. B. discussed the results and wrote the manuscript.

**Competing financial interests**

The authors declare no competing financial interests.

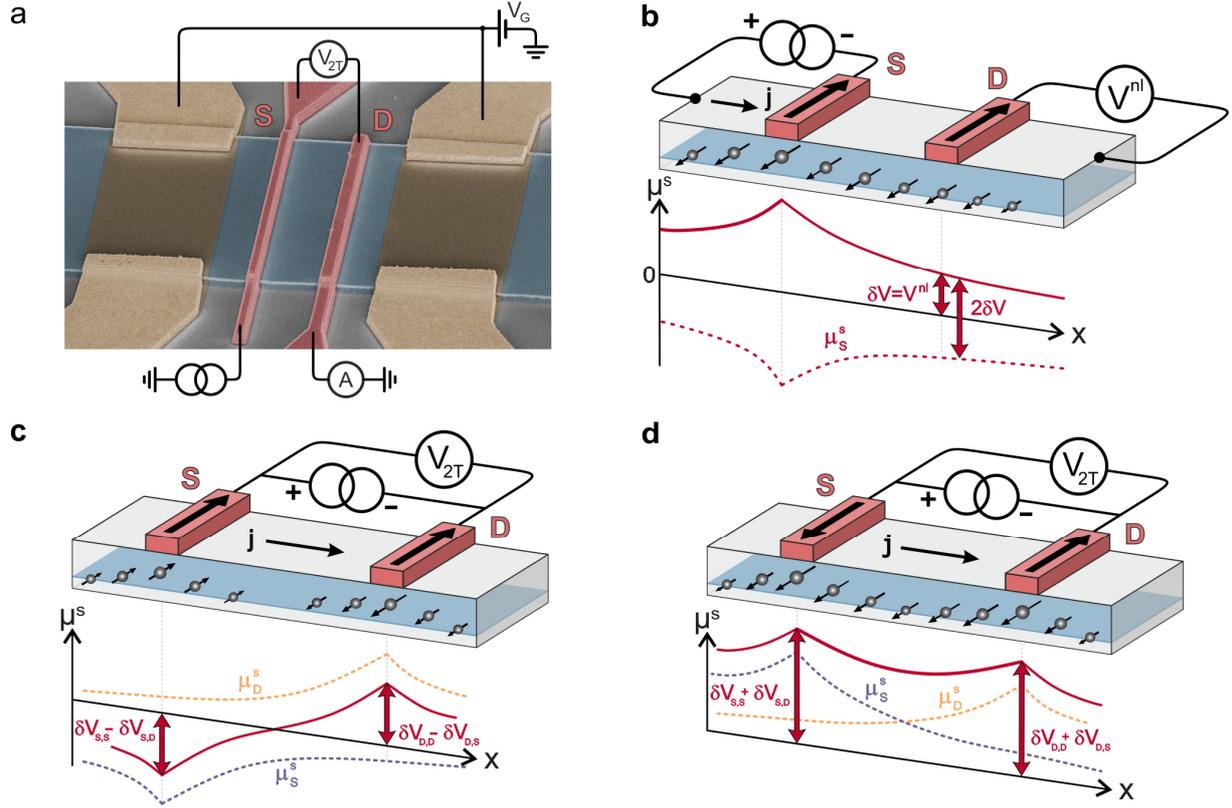

**Figure 1 | Devices and measurement configurations. a** Colored SEM image of one of the gated devices. Two narrow contacts in the middle, 500 nm and 700 nm wide, separated by 3.6 µm, are the ferromagnetic source (S) and drain (D) leads. Two electrostatic gates are used to confine the spins between S and D for $V_G < 0$. Additionally, the electric circuit for 2T measurements is sketched. **b** Nonlocal configuration, with S being an injector and D a detector. Majority spins are injected into the channel from the negatively biased S and diffuse along the channel. The solid line indicates the spin accumulation profile $\mu^s(x) = \frac{1}{2}(\mu_\uparrow(x) - \mu_\downarrow(x))$, where $\mu_{\uparrow(\downarrow)}$ is the quasichemical potential for the corresponding spin direction. The dashed line shows $\mu^s(x)$ when either the polarity of the injection current ***j*** or the magnetization direction of the source has been reversed. $\mu^s$ underneath the drain induces the spin-dependent voltage $\delta V$, measured nonlocally as $V^{nl}$, which changes by $\Delta V^{nl} = 2\delta V$ each time the magnetization of either of the contacts is reversed. **c, d** 2T

configuration with positively biased source and negatively biased drain for parallel and antiparallel orientation of the magnetization of the contacts, respectively. Solid lines indicate the total spin accumulation $\mu^s(x) = \mu_S^s(x) + \mu_D^s(x)$, where $\mu_{S(D)}^s$ (dashed line) is the spin accumulation generated at source (drain). In AP (P) configuration both components have the same (opposite) sign. As a result, $\mu^s$ is larger (smaller) in the AP (P) configuration. The 2T voltage difference between both configurations is given by $\Delta V = 2\delta V_{S,D} + 2\delta V_{D,S}$, where $\delta V_{i,j}$ is the voltage drop at contact *i* due to $\mu_j^s$.

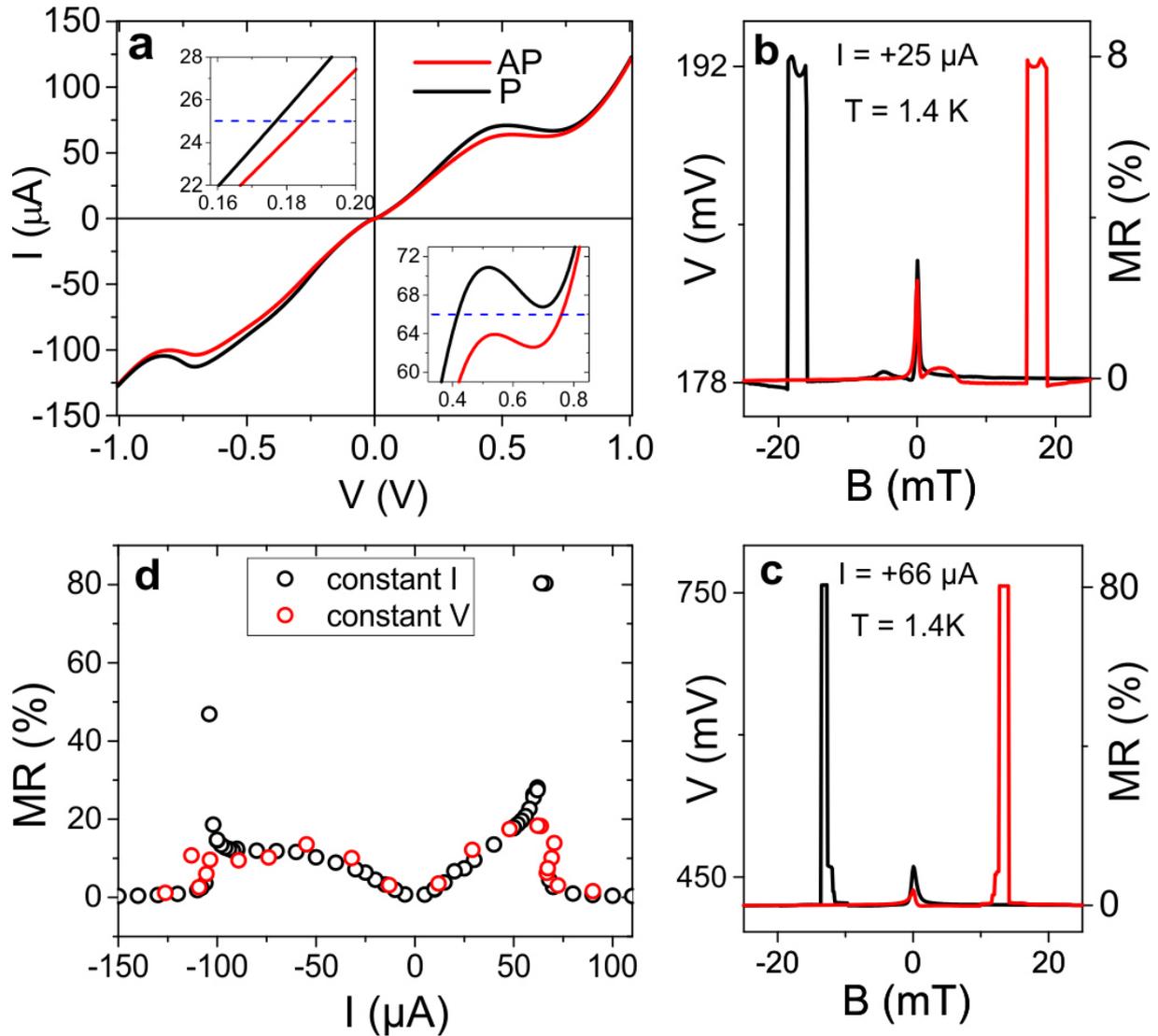

**Figure 2 | Magnetoresistance ratio and its bias dependence. a** *I-V* characteristics of the ungated device for P and AP orientation of source and drain magnetization. S–D separation *d*= 3.6 µm. Insets show sections of the *I-V* curve (same units as in main graph) in the linear regime at low bias (top) and in the nonlinear region of negative differential resistance (NDR) of the Esaki diode (bottom). Blue dashed lines indicate current values at which spin valve traces, shown in b and c, were taken. **b, c** Magnetoresistance of the device for ***B*** swept along the contacts' axes, taken at a constant current of *I*=+25 µA and +66 µA, respectively. Typical spin valve behavior is observed upon magnetization reversal with the higher voltage level corresponding to AP magnetization orientation of source and drain contacts. **d** Magnetoresistance ratio $MR = \Delta R/R^P$ as a function of the applied current. Results of both measurement

modes with constant current (O) and constant voltage (O) are shown. In the latter case, the MR is plotted versus the current in parallel configuration. Similar MR behavior was observed on 8 different devices.

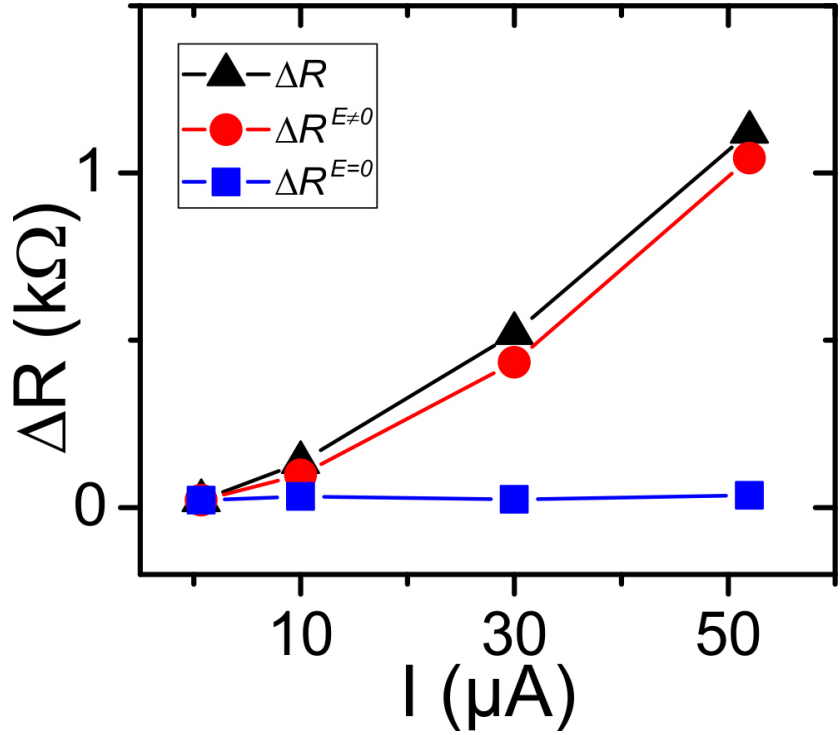

**Figure 3 | Influence of electric field on the magnetoresistance.** Comparison of the measured two-terminal spin signal $\Delta R$ (▲) with $\Delta R^{E=0} = \Delta R^{nl}_{S,D} + \Delta R^{nl}_{D,S}$ (■), obtained when neglecting electric field effects, and with $\Delta R^{E\neq0} = \eta_D \Delta R^{nl}_{S,D} + \eta_S \Delta R^{nl}_{D,S}$ (●), obtained when taking drift in the channel and electric fields in the contacts into account. Results are shown for four different currents between positively biased source and negatively biased drain. $\Delta R^{nl}_{S,D}$ and $\Delta R^{nl}_{D,S}$ were obtained from nonlocal measurements with unbiased detector and either positively biased source or negatively biased drain as a spin injector to match the polarity of the given lead in 2T measurements. Enhancement ($\eta_S$) and suppression ($\eta_D$) factors for positively biased S and negatively biased D, respectively, were obtained from nonlocal measurements with biased detectors (Supplementary Information 3). Solid lines are guides to the eye.

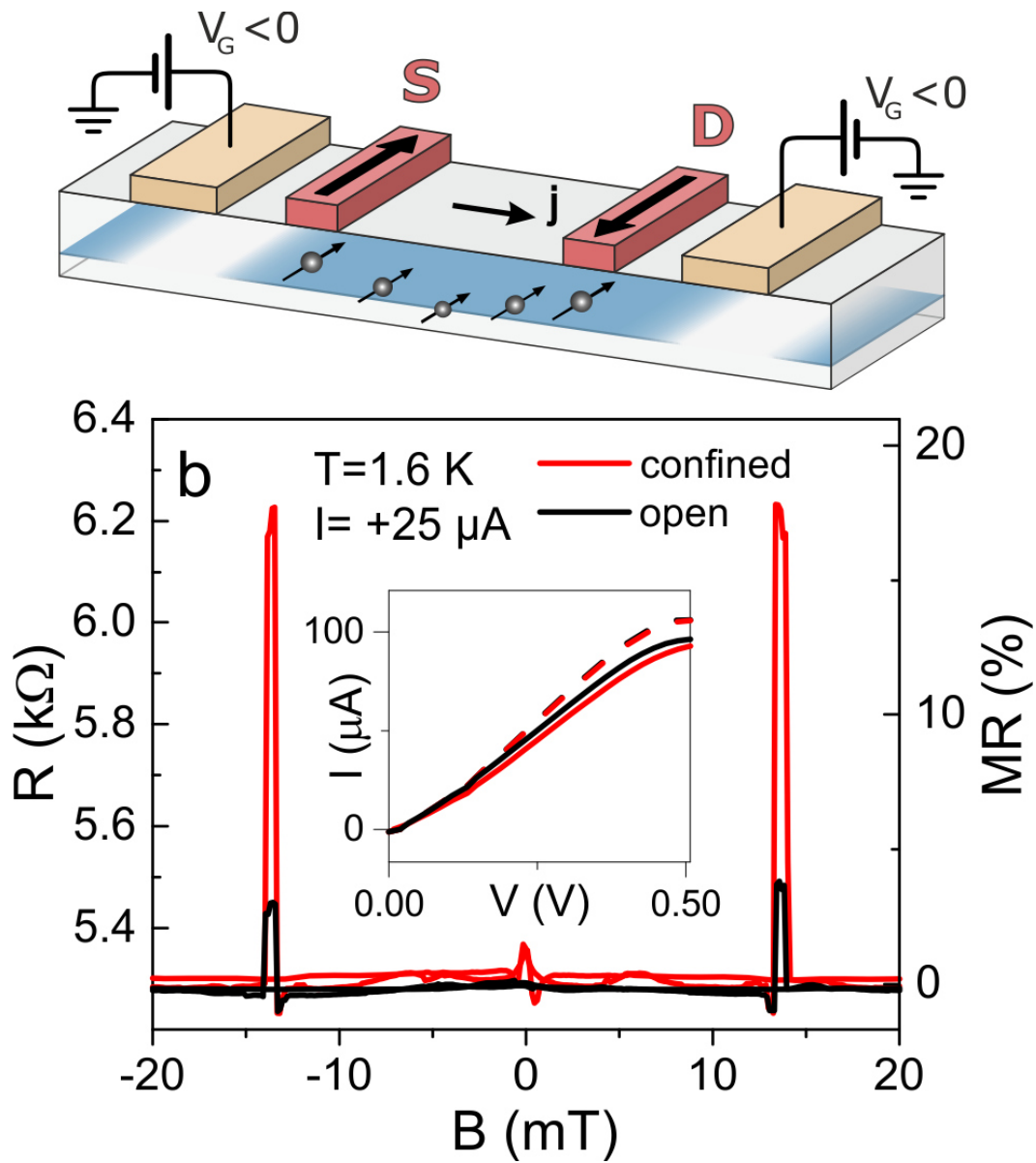

**Figure 4 | Tuning of the MR ratio with gate voltage. a** Sketch of the spin distribution for AP orientation of magnetization in the confined configuration, i.e., the 2DES outside the S and D contacts is depleted by applying a negative voltage $V_G$ to the gates. Spin accumulation between the contacts is increased, as spins cannot diffuse outside this region. **b** 2T SV traces with open channel (black line, $V_G$=+1.5 V) and confined channel (red line, $V_G$=-4 V). For $I$=−25 µA the resistance in the antiparallel configuration changes by 14%, corresponding to a voltage change of ~20 mV. Inset: part of the *I-V* curve in the corresponding bias range. Solid lines correspond to the AP configuration in the open (black) and the confined (red) case.

Dashed lines correspond to the P configuration in both cases. Similar measurements were performed on two different devices, giving comparable results (Supplementary Information 5).

# Supplementary information

**Content**

1. Device characterization.
2. Magnetoresistance of the gated device.
3. Influence of electric field on spin signal.
4. Spin diffusion calculation of ΔR in a confined geometry.
5. Gate control of the magnetoresistance.

**1. Device characterization.**

In this section, we present the results of basic characterization measurements of the non-gated spin injection device, used for the experiments summarized in Figure 2 and 3 of the main text.

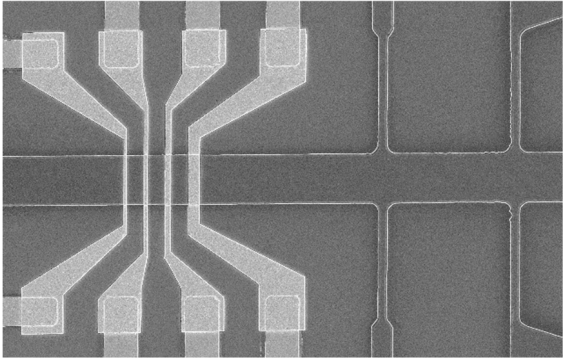

**Figure S1**. SEM picture of a typical non-gated device. It consists of a 10 μm wide channel with 4 ferromagnetic contacts (bright areas) on top. Four potential probes can be seen right of the magnetic contacts. Contacts to the mesa and to the voltage probes are not shown.

Figure S1 shows one of our devices without gate. The transport channel containing the 2DES is contacted by (Ga,Mn)As/GaAs Esaki diodes covered with a Au/Ti metal film which connects the degenerately p-doped (Ga,Mn)As to the contact pads, seen in the picture. The widths of the FM electrodes are 500 nm, 700 nm, 1 μm and 2 μm, separated by (center to center) 3.6 μm, 3.85 μm and 4.5 μm, respectively. The two narrowest electrodes were used as source

(S) and drain (D) in the two-terminal experiments described in the main text. The four potential probes were used to determine charge transport parameters in magnetotransport measurements (see Fig. S3). Six large 100 μm × 100 μm contacts are used to provide electrical contact to the channel: two at each end of the mesa and four at the voltage probes. They also consist of Esaki diodes topped with a Au/Ti layer. These contacts, far away (in terms of the spin diffusion length) from the spin-injector/detector contacts serve as current/voltage probes in magnetotransport measurements (Fig. S3) and as reference contacts in nonlocal spin injection experiments (Fig. S4).

Figure S2 shows the current-voltage (*I-V*) characteristics of an S and D contact. These traces are typical for our devices[1,2] employing Esaki tunnel diodes. They are nearly linear in the low-bias region up to $V_{3T} \cong \pm 0.25$ V. Here, $V_{3T}$ is the three-terminal voltage, measured as sketched in the inset of Fig. S2. Between 0.25 V ≲ $V_{3T}$ ≲ 0.6 V the non-linear region of negative differential resistance (NDR) prevails.

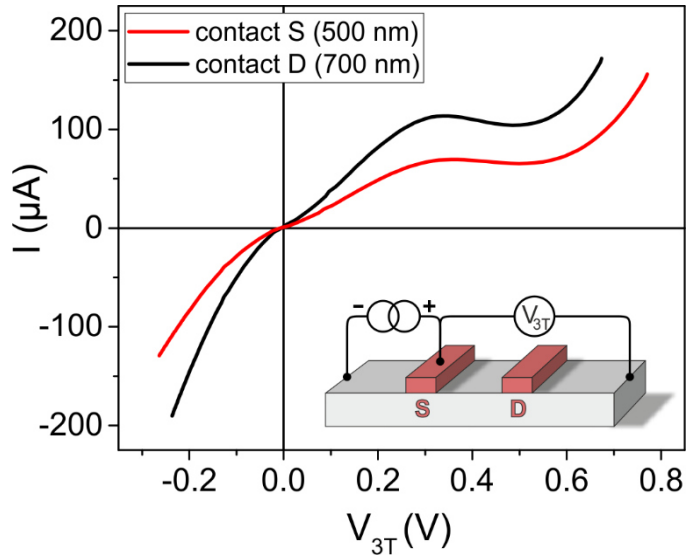

**Figure S2.** I-V characteristics of the two narrow FM electrodes, used as source and drain leads. $V_{3T}$ is the voltage drop across the interface, sketched in the inset. Zero-bias resistance area product is $R_{ZB}A \approx 2.5 \pm 0.3 \times 10^{-8}$ $\Omega m^2$

Figure S3 shows magnetotransport measurements performed in an out-of-plane magnetic field **B**. Shown are Hall resistivity $\rho_{xy}$ and longitudinal resistivity $\rho_{xx}$. From the measurements we extract the sheet resistance $R_s = 87\ \Omega$, carrier density $n_s = 2.5 \times 10^{11}$ cm$^2$, carrier mobility $\mu_e = 3.1 \times 10^5$ cm$^2$/Vs, and mean free path $l_e = 2.5$ μm.

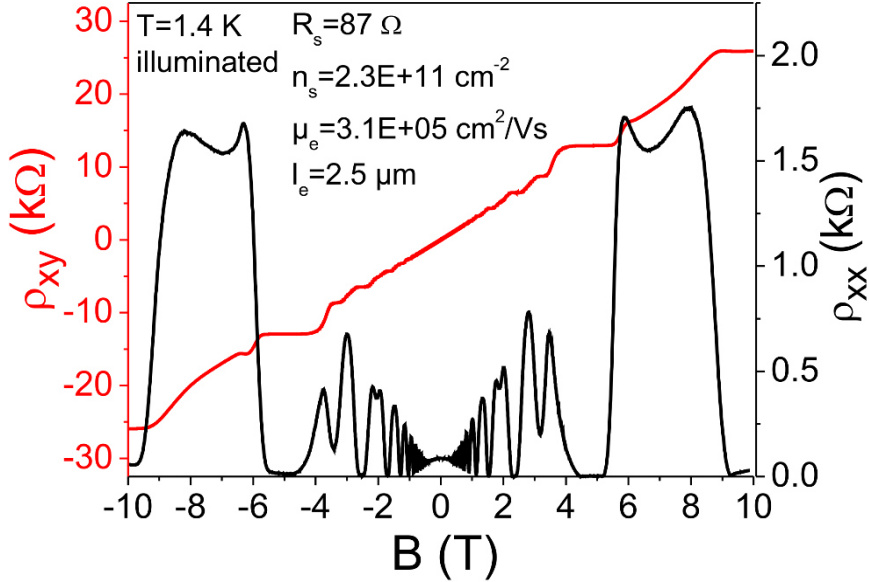

**Figure S3.** Hall resistivity ($\rho_{xy}$) and longitudinal resistivity ($\rho_{xx}$). The resistivity $\rho_{xx}$ vanishes at integer filling factors indicating that no parallel transport channel exists.

Figure S4 summarizes the results of nonlocal spin valve (NLSV) measurements performed with low AC excitation current, in the configuration as shown in Fig. S4b. Here, the 500 nm wide FM electrode serves as injector and the other contacts as nonlocal detectors. Spin accumulation generated at the injector diffuses along the channel and is detected by detectors as nonlocal voltage $V^{nl}$.[3] Sweeping **B** along the long axis of the FM contacts from +0.5 T to -0.5 T and back switches the detected $V^{nl}$ (Fig. S4a), whenever the magnetization orientation of injector and detector switches from parallel (P) to anti-parallel (AP) and vice versa. The amplitude of the NLSV signal, i.e. the change of the nonlocal voltage $\Delta V^{nl}$, is, in the tunneling regime of spin injection, given by[4]

$$\Delta V^{nl} = \frac{P_{inj} P_{det} I R_s \lambda_s}{W} \exp(-\frac{d}{\lambda_s}),$$

where $P_{inj(det)}$ is the spin injection (detection) sensitivity, $W$ the width of the channel, $R_s$ the sheet resistance, $\lambda_s$ the spin diffusion length and $d$ the injector-detector separation for a given pair of contacts.

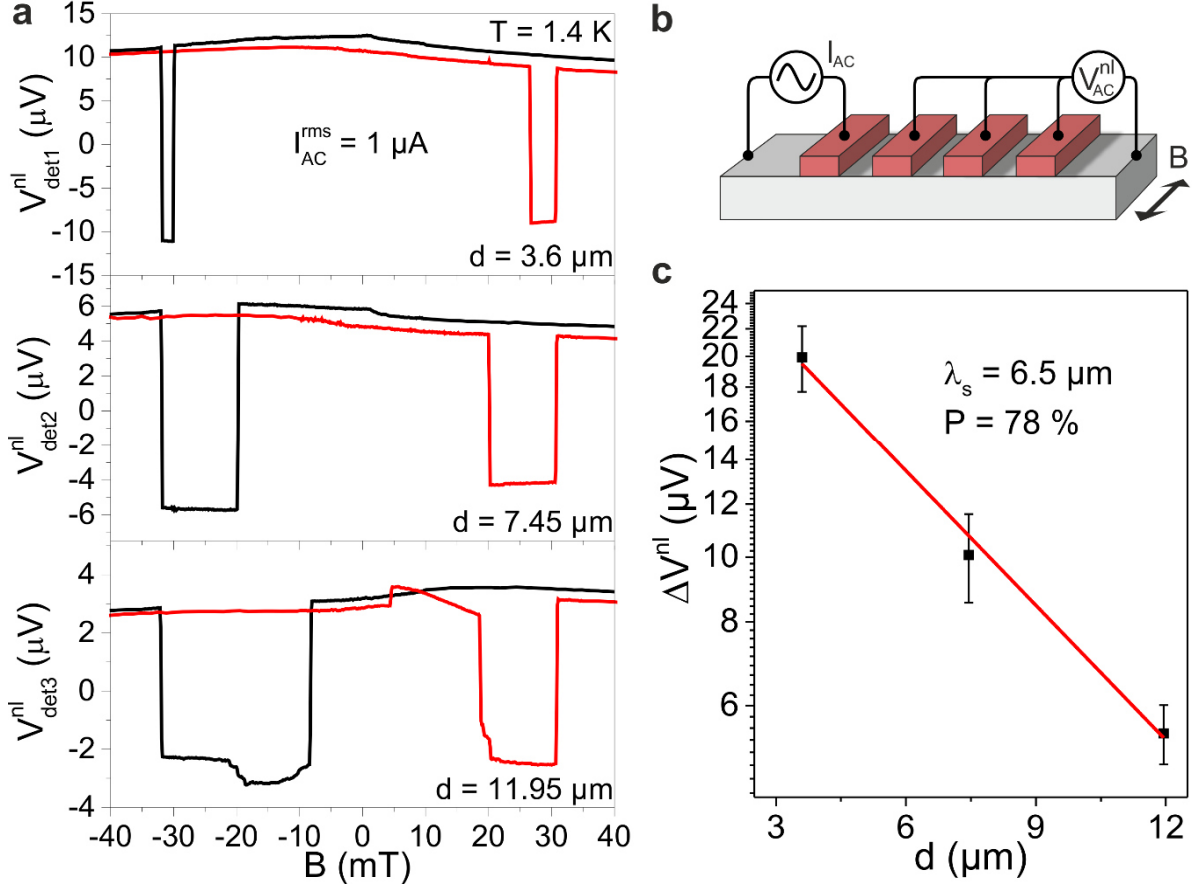

**Figure S4. (a)** Nonlocal spin valve experiments measured at three distances d from the injector using the configuration sketched in **(b)**. The AC (f=13 Hz) excitation current was $I_{AC}^{rms} = 1$ µA **(c)** Distance dependence of $\Delta V^{nl}$, used to extract the spin diffusion length $\lambda_s = 6.5$ µm and spin injection efficiency of P=78 %.

From the distance dependence $\Delta V^{nl}(d)$ (Fig. S4c) we extract the spin diffusion length $\lambda_s = 6.5$ µm, and the spin injection efficiency $P$=78%, assuming $P_{inj} \cong P_{det} = P$ for low bias values. Similar results were also obtained from low excitation DC measurements.

From the extracted charge transport and spin transport parameters we calculate the channel spin resistance as $R_{ch} = \rho \lambda_s \frac{1}{Wt} = R_s \lambda_s \frac{1}{W} \approx 57$ Ω, with the channel cross-section $A = Wt$, where $W = 10$ µm is the width and $t$ is the thickness of the 2DES channel.[5] The spin resistance of the tunnel contact we calculate as $R_T = \frac{R^{E=0}}{1-P^2} \approx 14.5$ kΩ. Both parameters determine the conditions for efficient spin injection and detection in a given system. According to the standard model, the maximum spin signal is observed when $R_{ch} \ll R_T \ll R_{ch} \lambda_s/d$.[5,6] The left condition is clearly satisfied for our device, indicating that the device works in the tunneling regime, enabling efficient spin injection. The right condition is not fulfilled, as the ratio $R_T/R_{ch} \approx 250$ is two orders of magnitude too large. This means that the dwell time of electrons in the channel is much larger than the spin relaxation time.[4–7] These values are typical for all our devices and a large $R_T/R_{ch}$ ratio is typical for semiconductor-based spin devices, resulting in a low magnetoresistance ratio MR $= \Delta R/R^P$.[8–13]

## 2. Magnetoresistance of the gated device.

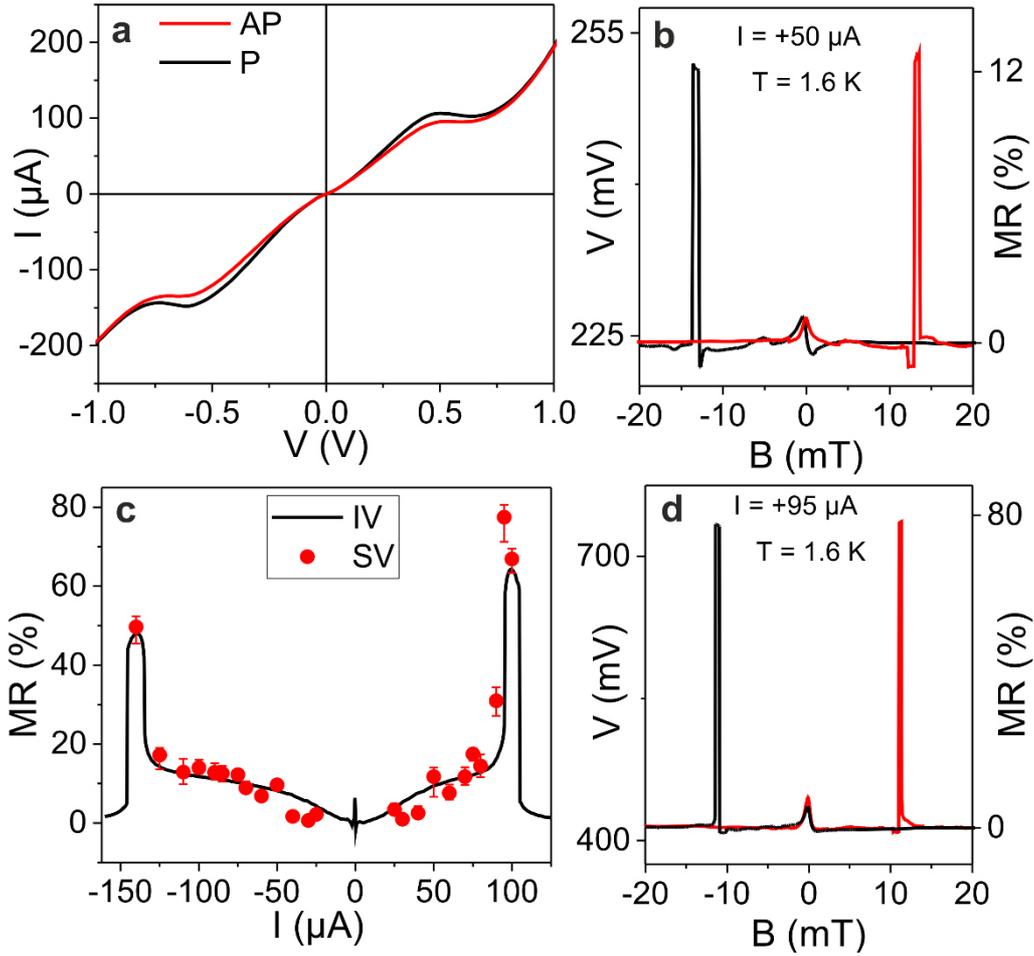

**Figure S5. a** Two-terminal I–V curves of the gated device in open configuration, with $V_G$=+1.5 V. **b,d** 2T spin valve measurements in the linear (**b**) and non-linear regime (**d**). **c** Bias dependence of the magnetoresistance ratio MR. Red circles indicate data points from spin valve measurements and the line represents data obtained by subtracting I–V curves obtained in the parallel and antiparallel configuration for each current.

In this section, we present additional two-terminal local measurements on the gated sample, used for the gate-controlled experiments presented in Figure 4 of the main text. The *I-V* characteristic measured between source and drain, shown in Fig. S5a, is almost point symmetric with respect to the origin. Spin valve signals are shown for two bias currents in Fig. S5b and d. The MR, i.e. $\Delta R/R^P$ with the resistance difference $\Delta R$ between parallel and anti-parallel orientation is shown in Fig. S5c as a function of bias current and reaches values of up to 80%.

## 3. Influence of an electric field on the spin signal.

In this section, we explain in detail how we constructed Figure 3 in the main text. There we compared the measured 2T local signal $\Delta R$ with values $\Delta R^{E=0}$ expected from standard theory without electric field effects, and with $\Delta R^{E\neq 0}$, for which electric field's effects have been included. As an example, we discuss the procedure for an injection current of 52 µA.

Fig. S6a displays the results of 2T local spin valve measurements with a positively biased source (S) contact. We measure a large SV signal $\Delta V = 60\ mV$, which we plot in Fig. 3 (black triangles, main text) as resistance $\Delta R = \frac{\Delta V}{I} = 1.1\ k\Omega$. This value we compare with $\Delta V^{E=0} = \Delta V_{S,D}^{E=0} + \Delta V_{D,S}^{E=0}$, where $\Delta V_{S,D}^{E=0}$ and $\Delta V_{D,S}^{E=0}$ correspond to, respectively, the voltage change at the source due to the spin accumulation generated at the drain and the other way around. Both $\Delta V_{S,D}^{E=0} = \Delta V_{S,D}^{nl}$ and $\Delta V_{D,S}^{E=0} = \Delta V_{D,S}^{nl}$ are measured in nonlocal experiments shown in figures S6b and S6c, respectively. From these experiments we get $\Delta V_{S,D}^{nl} = 0.83\ mV$ and $\Delta V_{D,S}^{nl} = 1.02\ mV$, resulting in $\Delta V^{E=0} = \Delta V_{S,D}^{nl} + \Delta V_{D,S}^{nl} = 1.85\ mV$. This and corresponding values taken at other bias currents are shown in Fig. 3 (blue squares) of the main text as resistance $\Delta R^{E=0} = \frac{\Delta V^{E=0}}{I}$. Next, we show how to estimate $\Delta V^{E\neq 0}$. We first decompose the signal as $\Delta V^{E\neq 0} = \eta_S \Delta V_{S,D}^{nl} + \eta_D \Delta V_{D,S}^{nl}$, where $\eta_{S(D)}$ is a factor representing the effect of the electric field on the signal component detected at the source (drain). We write this factor as $\eta_S = \eta_S^{(i)} \eta_S^{(ii)}$, where $\eta_S^{(i)}$ describes the effect of drift on spin transport in the channel[14] and $\eta_S^{(ii)}$ the effect of bias on the spin-to-charge conversion of the biased FM contact.[15,16]

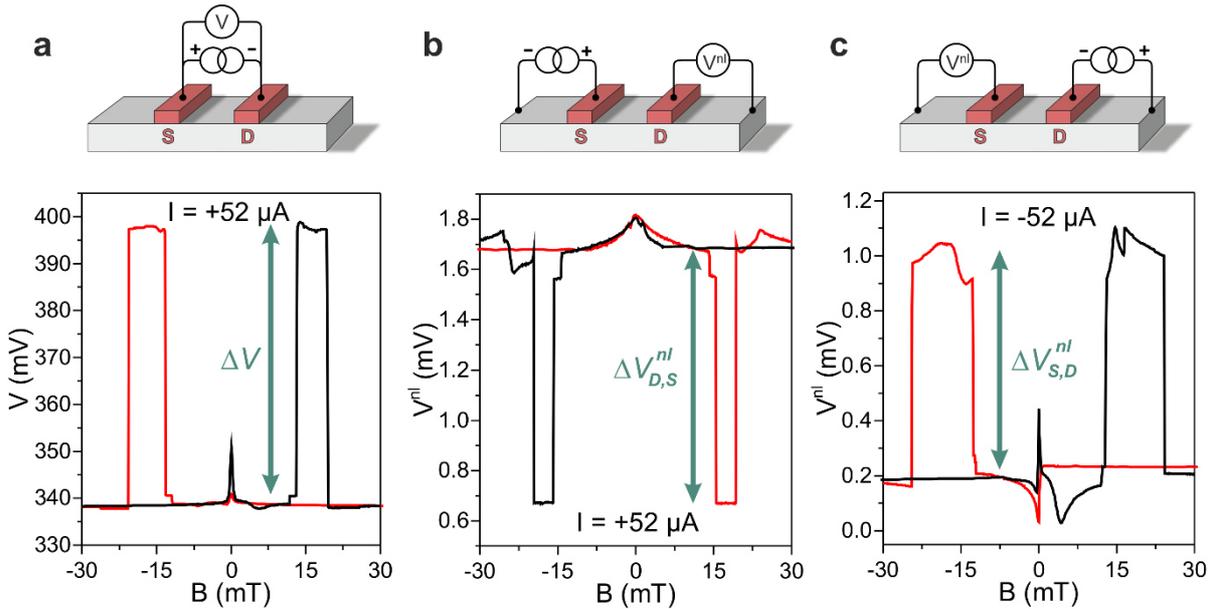

**Figure S6.** Two-terminal local (2TL) and nonlocal (NL) spin valve measurements for an injection current of 52 µA. **a** 2TL measurements with a positive bias on the source and a negative bias on the drain. **b** NL measurements with a positively biased S as an injector and the non-biased D as a detector. **c** NL measurements with a negatively biased D as an injector and non-biased S as a detector.

Estimate of $\eta_S^{(i)}$ and $\eta_D^{(i)}$: Drift in the channel influences the diffusion of spins along the channel, which in the absence of drift is described by the intrinsic spin diffusion length $\lambda_s$. The presence of an electric field enhances diffusion of spins in drift direction ("down-stream") and suppresses diffusion of spins in the opposite direction ("up-stream"). This can be described for both cases by a modified spin diffusion lengths, given by[14,17] $\lambda_{s,d(u)} = \left[ -(+) \frac{|E| \mu_e}{2 D_s} + \sqrt{\left(\frac{|E| \mu_e}{2 D_s}\right)^2 + \frac{1}{\lambda_s^2}} \right]^{-1}$, where $\lambda_s$ and $\lambda_{s,d(u)}$ are the intrinsic and down- (up-) stream spin diffusion lengths in the channel, respectively, $\mu_e$ is the electron mobility, $D_s$ is the spin diffusivity and $E$ the electric field. For 2T experiments as shown in Fig. S6a, i.e. with positively biased source and negatively biased drain, the drift carries electrons from drain towards source. Therefore, diffusion of spin accumulation from drain towards source is enhanced by drift and described by $\lambda_{s,d}$. The component of the spin valve signal $\Delta V_{S,D}$ due to spin accumulation generated at the drain and detected at the source is then enhanced by drift, with an enhancement factor defined as $\eta_S^{(i)} = \frac{\Delta V_{S,D}^{E\neq 0}}{\Delta V_{S,D}^{E=0}} = \frac{\exp(-\frac{d}{\lambda_{s,d}})}{\exp(-\frac{d}{\lambda_s})}$. Correspondingly, diffusion of spin accumulation from source towards drain is suppressed by drift and described by $\lambda_{s,u}$. The component of the spin valve signal $\Delta V_{D,S}$ due to spin accumulation generated at the source and detected at the drain is suppressed by drift, with a suppression factor defined as $\eta_D^{(i)} = \frac{\Delta V_{S,D}^{E\neq 0}}{\Delta V_{S,D}^{E=0}} = \frac{\exp(-\frac{d}{\lambda_{s,u}})}{\exp(-\frac{d}{\lambda_s})}$.

We now exemplarily calculate the factors $\eta_S^{(i)}$ and $\eta_D^{(i)}$ for $I$=52 µA. The corresponding electric field in the channel is $E = \frac{IR_S}{W} = 4.21$ V/cm ($W = 10$ µm is the channel width). $\lambda_s = 6.5$ µm and electron mobility $\mu_e$ are taken from measurements described in section 1. With the spin diffusivity $D_s = \lambda_s^2/\tau_s$, where the spin relaxation time $\tau_s \approx 1$ ns has been measured by Hanle spin precession[18,19] we obtain $\lambda_{s,d}(I = 52\ \mu A) = 16$ µm and $\lambda_{s,u}(I = 52\ \mu A) = 2.66$ µm from the expression given above. For the source–drain separation of 3.6 µm this gives the corresponding enhancement and suppression factors $\eta_S^{(i)}(I = 52\ \mu A) \cong 1.39$ and $\eta_D^{(i)}(I = 52\ \mu A) \cong 0.45$.

The effect of biased FM electrode on the measured spin signal, i.e. $\eta_S^{(ii)}$ and $\eta_D^{(ii)}$, we extract from the nonlocal experiments shown in Figures S7 and S8, respectively. Fig. S7a shows the nonlocal spin valve signal detected at the source contact with an AC injection current $I_{AC}^{rms} = 700$ nA flowing through the drain. The signal amplitude is $\Delta V_{S,D}^{nl} = 7.65$ µV. Fig. S7b displays the NLSV obtained when an additional DC current of $I_{DC}^S = +52$ µA is applied to the source, i.e. to the nonlocal detector. Note that both amplitude and polarity of the applied DC current match the magnitude and the polarity of the current flowing through the source in the 2T measurements, shown in Fig. S6a. The biased detector exhibits a much larger amplitude of the signal $\Delta V_{S,D}^{nl,E\neq 0}(I_{DC}^S = +52\ \mu A) = 337$ µV compared to the non-biased case of $\Delta V_{S,D}^{nl}(I_{DC}^S = 0) = 7.65$ µV. From both values we extract

$$\eta_S^{(ii)} = \Delta V_{S,D}^{nl,E\neq 0}(I_{DC}^S = +52\ \mu A)/\Delta V_{S,D}^{nl}(I_{DC}^S = 0) = 44.5,$$

i.e. biasing the source in positive direction enhances the signal detected at this contact by a factor of 44.5 for $I_{DC}^S = +52\ \mu A$.

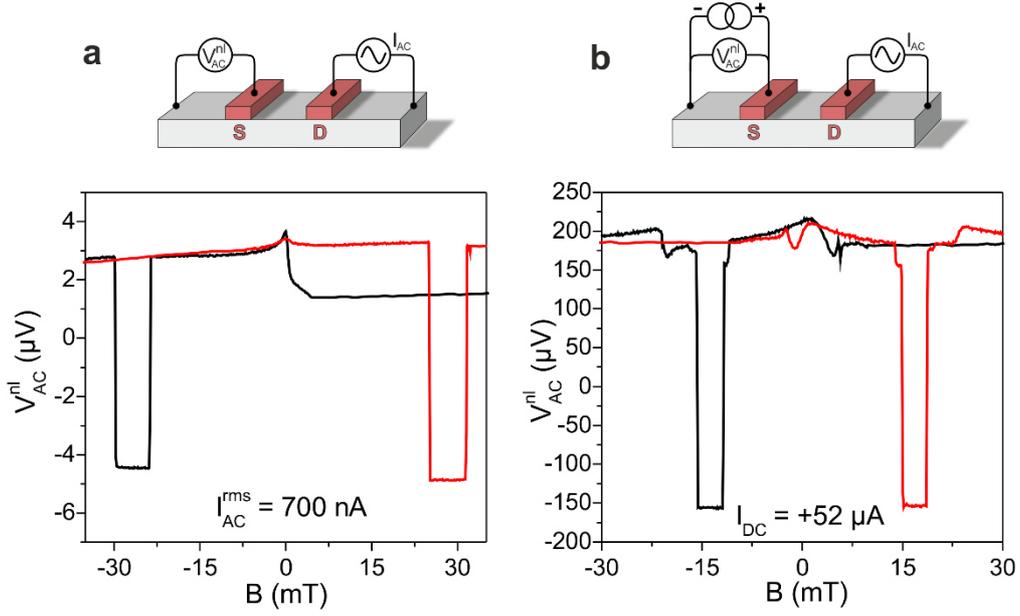

**Figure S7.** Measurements performed in order to determine the enhancement of the spin-related voltage detected at the source, as a result of biasing the lead in the forward direction.

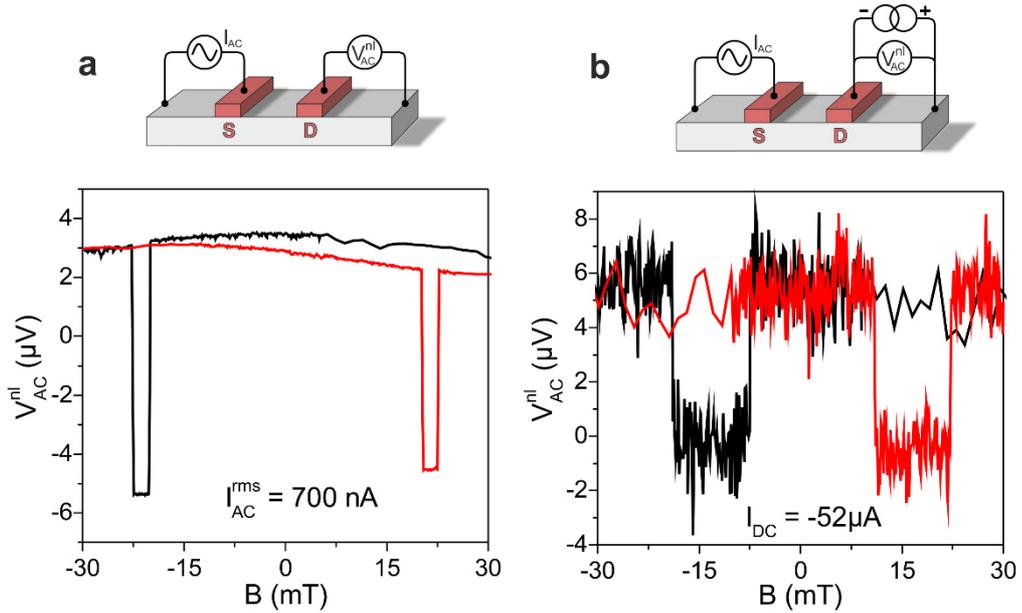

**Figure S8.** Measurements performed to determine the suppression of the spin-related voltage detected at the drain, as a result of biasing the lead in the reverse direction.

Fig. S8 shows analogous measurements with source as injector and drain as nonlocal detector for both non-biased drain (Fig. S8a) and drain biased with a current of $I_{DC} = -52$ μA (Fig. S8b). The negative bias applied to the drain suppresses somewhat the measured spin signal. The suppression factor $\eta_D^{(ii)}$ for $I_{DC}^D = -52$ μA is calculated as

$$\eta_D^{(ii)} = \Delta V_{D,S}^{nl,E\neq 0}(I_{DC}^D = -52\ \mu A)/\Delta V_{S,D}^{nl}(I_{DC}^{SD} = 0) = 0.76.$$

With $\eta_S(I = +52\ \mu A) = \eta_S^{(i)}\eta_S^{(ii)} = 61.8$ and $\eta_D(I = -52\ \mu A) = \eta_D^{(i)}\eta_D^{(ii)} = 0.34$ we calculate $\Delta V^{E\neq 0}(I = +52\ \mu A) = \eta_S \Delta V_{S,D}^{nl} + \eta_D \Delta V_{D,S}^{nl} = 54\ mV$, plotted in Fig. 3 of the main paper as $\Delta R^{E\neq 0}(I = +52\ \mu A) = \frac{\Delta V^{E\neq 0}}{I} = 1.04\ k\Omega$ as a red dot. In a similar way $\Delta R^{E\neq 0}$ values were obtained for other current values shown in Fig. 3. of the main text.

4. **Spin diffusion calculation of ΔR in a confined geometry.**

In this section we estimate the effect of confinement on the two-terminal spin resistance in the limit of low electric fields. For simplicity, we assume symmetric tunnel junctions having the same values for spin injection efficiency $P$ and contact resistance $R_T$. We checked that this assumption does not influence the following conclusions. As $R_F \ll R_{ch} \ll R_T$ holds for our devices (see Supplementary Information 1), we can estimate the 2T resistance change $\Delta R_{open}$ in open configuration from the standard diffusion equations:

$$\Delta R_{open} = 2P^2 R_{ch} \exp\left(-\frac{d}{\lambda_s}\right) \approx 40\ \Omega$$

For a confined geometry the formalism of Jaffrès et al.[5] predicts a two-terminal resistance of

$$\Delta R_{conf} = \frac{4P^2 R_T}{2\cosh\left(\frac{d}{\lambda_s}\right) + \left(\frac{R_T}{R_{ch}} + \frac{R_{ch}}{R_T}\right)\sinh\left(\frac{d}{\lambda_s}\right)} \approx 232\ \Omega$$

for the charge- and spin transport parameters of our device. This corresponds to an enhancement factor of $\Delta R_{conf}/\Delta R_{open} = 5.8$, which is in good agreement with our low field experimental results (see Fig. S9b).

## 5. Gate control of the magnetoresistance.

In this section we show the results of gate-controlled magnetoresistance measurements by repeating the measurement shown in Fig. 4 of the main text with another gated device. Fig. S9a shows *I-V* curves measured between source and drain contacts in both open and confined configurations. Similarly as for the sample discussed in the main text, there is a clear difference of the *I-V* trace for antiparallel magnetization orientation between open and closed configurations. This difference is reflected in the local spin valve measurements, shown in Fig. S9b, which displays the results for an injection current of I=32 µA. For this current we observe an increase of the MR ratio from ~3% in the open configuration to ~16% in the confined configuration, corresponding to a total resistance change of ~13%.

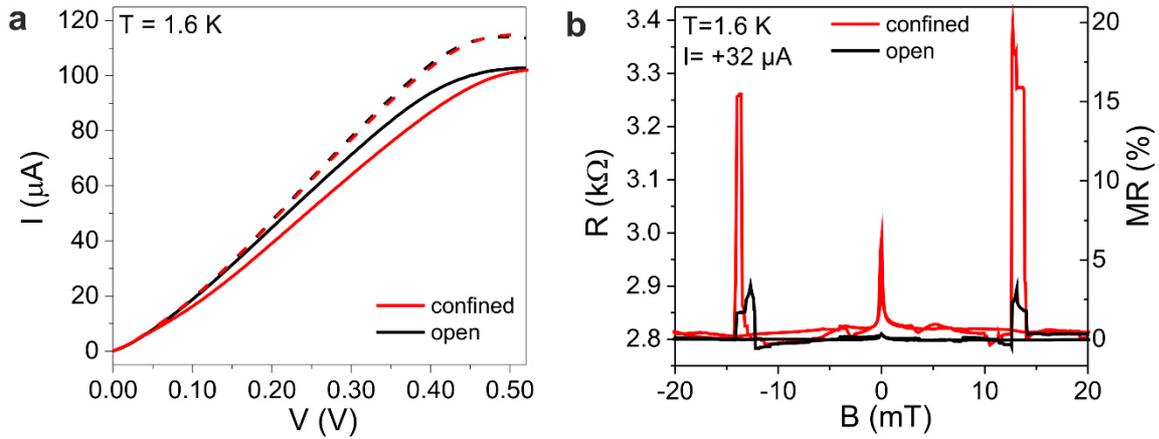

**Figure S9. a** I-V curve between source and drain contacts in open (black) and confined (red) configuration for parallel (dashed lines) and antiparallel (solid lines) orientation of the magnetization in both leads. **b** 2T local spin valve measurements in both open (black) and confined (red) configuration for an injection current I=32 µA. Measurements in the open configuration were performed with a gate voltage $V_G$ = +1.5 V and in the confined configuration with $V_G$ = -3 V.